# The Very Low Energy Solar Flux of Electron and Heavy-Flavor Neutrinos and Antineutrinos

W. C. Haxton

*Institute for Nuclear Theory, Box 351550, and Department of Physics*
*University of Washington, Seattle, Washington 98195-1550*
*email: haxton@phys.washington.edu*

Wei Lin[†]

*Center for Theoretical Physics and Department of Physics,*
*Massachusetts Institute of Technology, Cambridge, Massachusetts 02139-4307*
*email: wlin@iex.net*

(November 5, 2018)

## Abstract

We calculate the thermal flux of low-energy solar neutrinos and antineutrinos of all flavors arising from a variety of neutrino pair processes: Compton production (including plasmon-pole diagrams), neutral current decay of thermally populated nuclear states, plasmon decay, and electron transitions from free to atomic bound states. The resulting flux density per flavor is significant ($10^8 - 10^9$/cm$^2$/sec/MeV) below $\sim 5$ keV, and the distributions fill much of the valley between the high-energy edge of the cosmic background neutrino spectrum and the low energy tails of the pp-chain electron neutrino and terrestrial electron antineutrino spectra. Thermal neutrinos carry information on the solar core temperature distribution and on heavy flavor neutrino masses for $m_{\nu_\mu}$ or $m_{\nu_\tau} \gtrsim 1$ keV. The detection of these neutrinos is a daunting but interesting challenge.

A great deal of effort has been invested in detecting the solar electron neutrinos produced as a byproduct of solar fusion [1]. This letter focuses on another source of solar neutrinos, those produced by charged- and neutral-current thermal processes operating in the solar core. While the mean energy of these neutrinos is low, on the order of the solar core temperature $kT_c \sim 1.3$ keV, and their total flux quite modest, their flux density at earth is significant, ranging up to $\sim 10^9/\text{cm}^2/\text{sec}/\text{MeV}$ per flavor. Thus, above cosmic microwave energies $\epsilon_\nu \gtrsim 10^{-2}$ eV, this is the dominant source of low-energy heavy-flavor neutrinos, and of electron neutrinos below $\sim 5$ keV, where the low-energy tails of the solar fusion and terrestrial radioactivity neutrino spectra are encountered. In this letter we describe the processes producing such neutrinos and comment on the physics that might be extracted from their flux.

The pair neutrino mechanisms we consider are illustrated in Fig. 1. The first is Compton production, a mechanism we find dominates the high-energy tail of the thermal distribution. The second is Compton production with a plasmon intermediate state, a process we believe has not been considered previously. The third is the decay of a plasmon into a $\nu\bar{\nu}$ pair. Both the ordinary Compton and plasmon processes are familiar neutrino cooling processes in stars [2]. The fourth is the decay of a thermally populated excited nuclear state by pair emission, a process that can contribute at solar temperatures only in the case of an M1 transition at an anomalously low energy. Bahcall, Treiman, and Zee [3] identified the 14.4 keV Mossbauer transition in $^{57}$Fe as the principal contributor in the sun. The fifth process is pair emission as an electron makes the transition from a free state to a bound atomic orbital, the analog of an electromagnetic process known to be quite important to the solar opacity [4].

The Compton process, as well as the other three leptonic mechanisms mentioned above, are governed by the weak Hamiltonian

$$H = -\frac{G}{2\sqrt{2}}\bar{e}(g_V\gamma_\mu - g_A\gamma_\mu\gamma_5)e\bar{\nu}\gamma^\mu(1-\gamma_5)\nu. \tag{1}$$

The couplings for heavy flavor neutrinos are $g_V = 1 - 4\sin^2\theta_w$ and $g_A = 1$, while for electron neutrinos one obtains, after a Fierz transformation of the charged current interaction, the effective couplings $g_V = -(1 + 4\sin^2\theta_w)$ and $g_A = -1$. The Compton rate per unit volume is then given by

$$\frac{dN_C}{dt} = \int n_\gamma(k)\frac{d^3k}{2\omega(2\pi)^3}n_e(p)\frac{m_e d^3p}{\omega_p(2\pi)^3}(2\pi)^4\delta^{(4)}(p+k-p'-q)$$
$$\sum_{\{m\}}|M_C|^2(1-n_e(p'))\frac{m_e d^3p'}{\omega_{p'}(2\pi)^3}\frac{d^3p_\nu}{2\epsilon_\nu(2\pi)^3}\frac{d^3p_{\bar{\nu}}}{2\epsilon_{\bar{\nu}}(2\pi)^3}, \tag{2}$$

where $(\omega, \vec{k})$ is the four-momentum of the incident photon, $q = (\omega_\gamma, \vec{q}) \equiv p_\nu + p_{\bar{\nu}}$, $\omega_p = m_e + \epsilon_p \approx m_e + \vec{p}^2/2m_e$, and $\sum_{\{m\}}$ denotes the summation over the spins and polarizations. The Fermi and Bose-Einstein thermal distribution functions are

$$n_e(p) = \frac{1}{\exp[(\epsilon_p - \mu)/T] + 1}$$
$$n_\gamma(k) = \frac{1}{\exp[\omega_\gamma/T] - 1}. \tag{3}$$

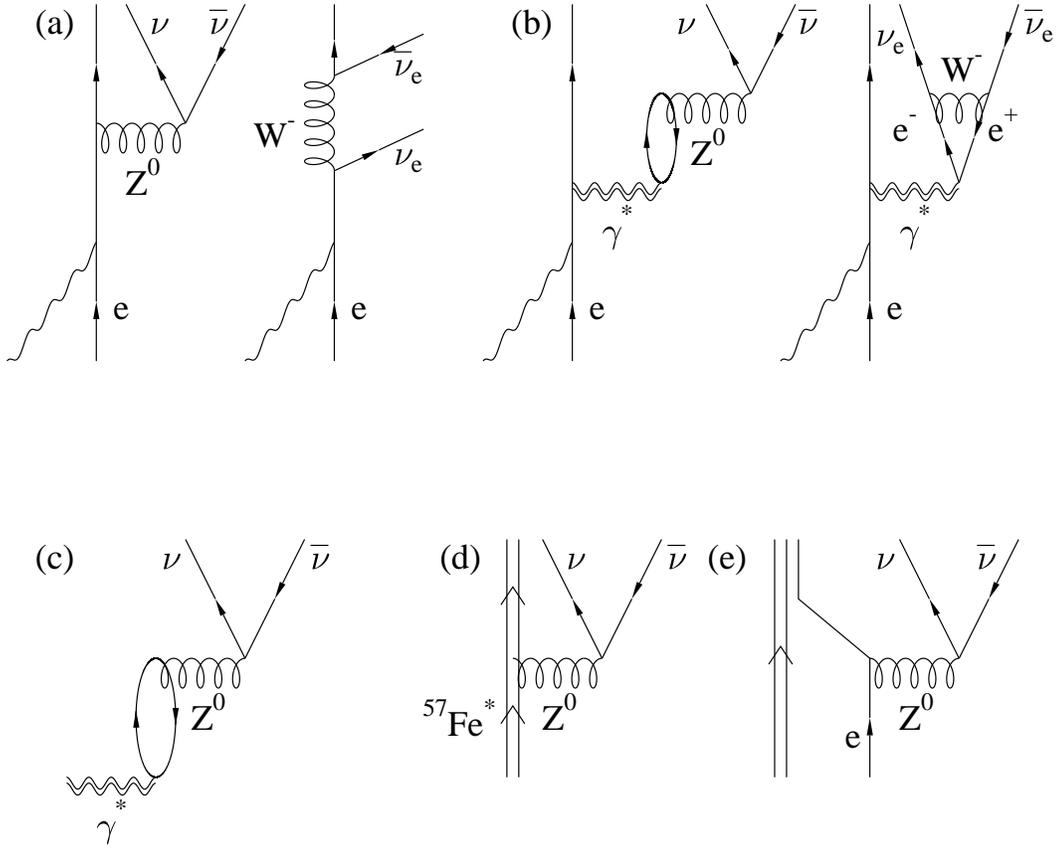

FIG. 1. Representative diagrams for the various thermal neutrino pair processes considered here: a) Compton process; b) plasmon pole contribution to the Compton process; c) transverse plasmon decay; d) nuclear $Z^0$ emission; and e) pair production in free-bound atomic transitions.

The chemical potential $\mu$, which is determined by the electron density and temperature, is $\sim -2.1$ keV at the solar core, so that the electron gas is slightly degenerate.

The spin sum yields

$$\sum_{\{m\}} |M_C|^2 = \frac{\pi\alpha G^2}{2m_e^2\omega^2} A^{\mu\nu} B_{\mu\nu}, \qquad (4)$$

where the neutrino tensor is

$$A^{\mu\nu} = 8[(p_\nu^\mu p_{\bar\nu}^\nu + p_\nu^\nu p_{\bar\nu}^\mu - p_\nu \cdot p_{\bar\nu} g^{\mu\nu}) + i\epsilon^{\rho\mu\sigma\nu} p_{\nu\rho} p_{\bar\nu\sigma}]. \qquad (5)$$

After making a nonrelativistic reduction the electron tensor is found to be

$$B_{\mu\nu} = \sum_{\{\epsilon\}} \left\{ 2(g_V^2 + g_A^2)(-\omega\epsilon_\mu + \delta_{\mu 0}\epsilon \cdot q)(-\omega\epsilon_\nu + \delta_{\nu 0}\epsilon \cdot q) + 2g_A^2[k_\mu k_\nu - (\epsilon \cdot q)^2 g_{\mu\nu}] \right\}. \qquad (6)$$

The remaining evaluation is simplified by the fact that $T \ll m_e$, allowing one to treat the electrons in a static approximation. They absorb momentum but no energy, $\omega \sim \omega_\gamma \equiv \epsilon_\nu + \epsilon_{\bar\nu}$, which leads to

$$\frac{dN_C}{dt} = \frac{n'_e G^2 \alpha}{6\pi^4 m_e^2}(g_V^2 + 5g_A^2) \int_0^\infty d\epsilon_\nu d\epsilon_{\bar\nu} n_\gamma \frac{\epsilon_\nu^2 \epsilon_{\bar\nu}^2}{\omega_\gamma}\left(\omega_\gamma^2 - \frac{2}{3}\epsilon_\nu \epsilon_{\bar\nu}\right), \tag{7}$$

where $n'_e = 2/(2\pi)^3 \int d^3p\, n_e(p)(1 - n_e(p))$ .

The plasma produces an interesting modification of this result, illustrated in Fig. 1b. The long-range Coulomb force between charges in the plasma produces a collective plasma oscillation, the longitudinal plasmon, characterised by the plasma frequency $\omega_{pl}$. The transverse plasmons are photon modes that develop an effective mass equal to $\omega_{pl}$ ($\hbar = c = k_B = 1$).

Neglecting the small contribution due to ionic charges, the plasma frequency is given by $\omega_{pl}^2 = \frac{4\pi n_e \alpha}{m_e}$, where $\alpha$ is fine-structure constant, $m_e$ is the electron mass, and $n_e = N_A \rho/\mu_e$ is the electron number density. Here $N_A$ is Avogadro's number, $\mu_e = 2/(1 + X)$ is the electron's effective molecular weight, and $X$ is the mass fraction of hydrogen. At the center of the sun, $\omega_{pl} \sim 0.28$ keV. In the classical limit the dispersion relation for the transverse plasmon is

$$\omega_\gamma^2 = \omega_{pl}^2(1 + \vec{q}^2/k_D^2) + \vec{q}^2, \tag{8}$$

while that for the longitudinal plasmon is $(0 \leq |\vec{q}| \lesssim \omega_{pl})$

$$\omega_\gamma^2 = \omega_{pl}^2(1 + 3\vec{q}^2/k_D^2). \tag{9}$$

Here $\omega_\gamma$ is the energy of the plasmon and $\vec{q}$ is the momentum. Since the Debye wavenumber $k_D = \sqrt{\frac{4\pi\alpha n_e}{T}}$ is much greater than $\omega_{pl}$, the plasmon energy is always very close to the plasma frequency.

The plasmon contributions to the Compton process involve both the vector–vector polarization $\Pi_{\mu\nu}^{VV}$ and the vector–axialvector polarization $\Pi_{\mu\nu}^{VA}$. The terms involving $\Pi_{\mu\nu}^{VA}$ are negligible, suppressed by $(\alpha \frac{T}{m_e})^2$ in the rate. While the simple Compton and pole amplitudes interfere, this interference can be ignored since the pole contribution is so narrowly peaked at kinematics defined by $q_\mu^2 \sim \omega_{pl}^2$. It follows that the pole contribution simply *adds* to Eq. (6) the following term

$$B_{\mu\nu}^P = 2g_V^2[(\varepsilon^{-1})_{\mu\sigma}(\varepsilon^{-1})^*_{\nu\rho} - g_{\mu\sigma}g_{\rho\nu}]\sum_{\{\epsilon\}}(-\omega\epsilon^\sigma + \delta_{\sigma 0}\epsilon \cdot q)(-\omega\epsilon^\rho + \delta_{\rho 0}\epsilon \cdot q), \tag{10}$$

where $(\varepsilon^{-1})_{\mu\nu} = g_{\mu\nu} + D_{\mu\sigma}\Pi^\sigma_{\ \nu}$ is the inverse of the usual dielectric tensor of the plasma, $D_{\mu\nu}$ being the full photon propagator. (The zeros of the determinant of the dielectric tensor determine the plasmon modes.) As this contribution is proportional to $g_V^2$, it is numerically significant only for electron neutrinos.

The net contribution of the plasmon pole can be evaluated by integrating over a sharply-peaked off-shell region determined by the small width of the plasmon. The plasmon width is zero in the random phase approximation (decay into one-particle one-hole electron states is forbidden for timelike $q_\mu^2$). Thus one must go to the next order: we take the width from the two-particle two-hole calculation of DuBois *et al* [5]. As we are primarily interested in neutrino energies above $\omega_{pl}$, the longitudinal mode can be ignored. A somewhat tedious calculation then yields the additional contribution to the Compton rate from transverse plasmons

$$\frac{dN_P}{dt} = \frac{n'_e G^2 \alpha}{48\pi^4 m_e^2} g_V^2 \int d\epsilon_\nu d\epsilon_{\bar{\nu}} n_\gamma \frac{\omega_{pl}^6 \pi}{\omega_\gamma^3 \mathrm{Im}\epsilon_t} (\epsilon_\nu^2 + \epsilon_{\bar{\nu}}^2). \tag{11}$$

The imaginary part of the transverse dielectric function is [5]

$$\mathrm{Im}\epsilon_t = \frac{\lambda}{10\pi^2} \frac{\vec{q}^{\,2}}{k_D^2} \frac{\omega_{pl}^5}{\omega_\gamma^5} [4J(\omega_\gamma)/3 + I(\omega_\gamma)], \tag{12}$$

where the plasma parameter $\lambda = k_D^3/n_e$, and

$$\begin{aligned} J(\omega_\gamma) &\sim 2\sqrt{\pi}, \\ I(\omega_\gamma) &\sim \sqrt{\pi} \int_0^\infty \frac{dx}{2x} \exp[-x - \omega_\gamma^2/(16T^2 x)]. \end{aligned} \tag{13}$$

These expressions are valid for $\omega_\gamma \gg \omega_{pl}$ and $|\vec{q}| \ll k_D \sim 5.5$ keV. The lower bound on $\omega_\gamma$ is important: plasmon contributions are large at low energies, so that later numerical results should be viewed with caution in this region. The constraint for large $|\vec{q}|$ is not a concern. At large $|\vec{q}|$ and thus large $\epsilon_\nu + \epsilon_{\bar{\nu}}$, the pole contribution is small and thus uninteresting relative to other processes: the pole diagrams become $O(\alpha)$ corrections to the vector-coupling part of the simple Compton process.

We note that the analogous Compton plasmon pole calculation for very hot or dense stars would require a finite-temperature relativistic calculation for the imaginary part of the polarization insertions beyond the random phase approximation. We believe no such calculation has yet been performed.

The contribution of direct transverse plasmon decays into $\nu\bar{\nu}$ is illustrated in Fig. 1c. Note that this process and the Compton plasmon pole process are entirely distinct: the plasmon is off shell in the pole process. The decay rate per unit volume is given by

$$\frac{dN_\gamma}{dt} = \int n_\gamma \frac{d^3 q}{2\omega_\gamma (2\pi)^3} \frac{d^3 p_\nu}{2\epsilon_\nu (2\pi)^3} \frac{d^3 p_{\bar{\nu}}}{2\epsilon_{\bar{\nu}} (2\pi)^3} (2\pi)^4 \delta^{(4)}(p_\nu + p_{\bar{\nu}} - q) \sum_{\{\epsilon\}} |M_\gamma|^2 , \tag{14}$$

where $\sum_{\{\epsilon\}}$ denotes the summation over the two polarizations and

$$|M_\gamma|^2 = A^{\mu\nu} (\Gamma_{\mu\alpha} \epsilon^\alpha)(\Gamma_{\nu\beta} \epsilon^\beta)^* , \tag{15}$$

with $\Gamma_{\mu\alpha}$ the effective photon-neutrino coupling [6] and $\epsilon_\alpha$ the photon polarization four-vector. The neutrino tensor $A^{\mu\nu}$ is given by Eq. (5), while $\Gamma^{\mu\alpha}$ can be determined from the VV and VA polarization insertions. The latter is again negligibly small. We find

$$\Gamma^{\mu\alpha} \epsilon_\alpha = -\frac{G}{4\sqrt{2\pi\alpha}} g_V \Pi_t(q)(0, \vec{\epsilon})^\mu , \tag{16}$$

where the transverse VV polarization insertion can be approximated by $\Pi_t(q) \sim \omega_{pl}^2$. The final result then reduces to

$$\frac{dN}{dt} = \frac{G^2}{128\pi^4} \frac{g_V^2}{\alpha} \omega_{pl}^6 \int \frac{1}{\omega_\gamma^2} \frac{1}{e^{\omega_\gamma/kT} - 1} (\epsilon_\nu^2 + \epsilon_{\bar{\nu}}^2) d\epsilon_\nu d\epsilon_{\bar{\nu}}, \tag{17}$$

where $\omega_\gamma = \epsilon_\nu + \epsilon_{\bar\nu}$ and the integrations extend over all neutrino energies satisfying $\omega_\gamma > \omega_{p\ell}$. We have ignored the contribution of longitudinal plasmons, which again contribute only if $\epsilon_\nu + \epsilon_{\bar\nu} \sim \omega_{p\ell}$.

Neutral current decays of excited nuclear states can produce pair neutrinos (Fig. 1d). The importance of a given element depends on the probability that the excited state will be thermally populated, on the element's solar mass fraction, and on the strength of the isovector axial (M1) transition matrix element. As noted in Ref. [3], the important metal is $^{57}$Fe because of the low-lying $3/2^- \to 1/2^-$ Mössbauer transition, despite the low solar abundance of this isotope (mass fraction of $3.26 \cdot 10^{-5}$ [7]). We can express [8] the pair neutrino rate per flavor in terms of the known gamma decay rate $\omega_\gamma = 7.41 \cdot 10^5$/sec (thereby eliminating most of the nuclear physics uncertainty)

$$\frac{dN}{dt} = \frac{\omega_\gamma}{1+\delta^2}\, N_{57}\, \frac{3}{\pi^3}\, \frac{G^2 F_A^2}{\alpha}\, \frac{M_N^2}{W_0^3}\, \frac{e^{-W_0/T}}{[\mu_1 + (\mu_0 - \frac{1}{2})\beta - \eta]^2} \int_0^{W_0} \epsilon_\nu^2 (W_0 - \epsilon_\nu)^2\, d\epsilon_\nu, \qquad (18)$$

where $N_{57}$ is the number density of $^{57}$Fe nuclei in the solar core, $W_0 = 14.4$ keV is the $3/2^- \to 1/2^-$ transition energy, $\delta = 0.0022$ is the E2/M1 mixing ratio, $\mu_0 = 0.88$ and $\mu_1 = 4.71$ are the isoscalar and isovector magnetic moments, and $F_A = 1.26$ is the axial vector coupling constant. The nuclear-structure-dependent terms

$$\eta = -\frac{<J_f||\sum_{i=1}^A \vec{\ell}(i)\tau_3(i)||J_i>}{<J_f||\sum_{i=1}^A \vec{\sigma}(i)\tau_3(i)||J_i>} \sim 0.80$$

$$\beta = \frac{<J_f||\sum_{i=1}^A \vec{\sigma}(i)||J_i>}{<J_f||\sum_{i=1}^A \vec{\sigma}(i)\tau_3(i)||J_i>} \sim -1.19 \qquad (19)$$

are taken from the shell model calculations of Ref. [8]. Note that the maximum rate for Eq. (18) occurs when $W_0 = 5T$, or about 6.5 keV in the solar core, so the $^{57}$Fe transition would be more effective in somewhat hotter stars.

The final neutrino pair process we consider is $Z^0$ emission in the transition of a continuum electron to a bound atomic orbital (Fig. 1e). Although such free-bound transitions to atomic states in Fe and other metals are known to be an important contributor to the solar opacity [4], relatively little attention has been paid to the analogous weak process. Early work was flawed by the use of plane waves for the continuum state: the resulting nonorthogonality of the initial and final wave functions leads to an overestimation of the rate by a factor proportional to $m_e/T$. More recently the process was treated correctly by Kohyama, Itoh, Obama, and Mutoh [9], who tabulated total cooling rates for a variety of metals using the dipole approximation (which we also adopt) and the transition to the $n=1$ $\ell=0$ atomic bound state.

The rate for capturing into Bohr orbits described by $(n, \ell)$ is

$$\frac{dN}{dt} = N(Z)\, f(n, \ell) \int d^3 \vec{p}\, \frac{2}{(2\pi)^3 (e^{(\epsilon_p - \mu)/kT} + 1)}\, v_p\, \sigma(p) \qquad (20)$$

where $\epsilon_p = \epsilon_\nu + \epsilon_{\bar\nu} - E_b(n, \ell)$ is the continuum electron's energy, $E_b(n, \ell)$ is the (positive) Bohr orbit binding energy, $N(Z)$ is the number density of the atoms in the solar core, $v_p$ is the electron velocity, $\sigma(p)$ is the cross section for capture of an electron of momentum $p$ on the ion (averaged over initial electron spin and summed over all final lepton spins), and $f(n, \ell)$ is a blocking factor correcting for the incomplete ionization of the atomic shells. (In the solar interior, the only atomic shells not completely ionized are those with low $n$ and high $Z$.)

The differential cross section can be readily evaluated in the dipole approximation, as discussed in Ref. [9]. The result is

$$\sigma(p) = \frac{1}{v_p} \frac{G^2}{9\pi^2} q_0^7 \left( \ell |\langle n\ \ell |r| p\ \ell - 1 \rangle|^2 + (\ell + 1)|\langle n\ \ell |r| p\ \ell + 1 \rangle|^2 \right)$$

$$\times \int_0^1 dx (1-x)^2 x^2 \left[ g_V^2(2x^2 - 2x + 3) + 2g_A^2(5x^2 - 5x + 3) \right.$$

$$\left. \pm \frac{q_0}{m_e} g_V g_A (16x^3 - 24x^2 + 14x - 3) + (\frac{q_0}{2m_e})^2 g_V^2 (28x^4 - 56x^3 + 46x^2 - 18x + 3) \right] \quad (21)$$

where $q_0 = \epsilon_\nu + E_b(n, \ell) = \epsilon_\nu + \epsilon_{\bar{\nu}}$, $x = \epsilon_\nu/q_0$, and where the radial integrals are defined with unit normalization for the incoming continuum electron wave function of momentum $p$. The $\pm$ sign on the third term distinguishes the $\nu$ (+) and $\bar{\nu}$ (-) spectral distributions. The radial integrals between hydrogenic bound states and spherical components of the incoming Coulomb wave of momentum $\vec{p}$ can be evaluated in closed form [10]. The integration over the neutrino spectrum can be readily performed to give

$$\sigma(p) = \frac{1}{v_p} \frac{G^2}{105\pi^2} q_0^7 \left( g_V^2 + \frac{3}{2} g_A^2 + \frac{5}{27} g_V^2 (\frac{q_0}{2m_e})^2 \right) M^2, \quad (22)$$

where $M^2$ is the matrix element of Eq. (21). The first two terms agree with the corresponding calculation of Ref. [9]. In using Eq. (21) to evaluate the free-bound neutrino spectrum, we ignore the terms proportional to $q_0/m_e$, which are clearly quite small at solar temperatures. This eliminates the VA interference term that distinguishes the $\nu$ and $\bar{\nu}$ distributions.

The qualitative features of free-bound transitions can be understood readily. As the four-momentum transfer to the neutrinos is timelike, the corresponding momentum transfer $|\vec{q}| \leq q_0 \sim \frac{3}{2} T \sim 2$ keV, so that $|\vec{q}| a_0/Z \sim 0.5/Z$, where $a_0/Z$ is the Bohr radius of the element of charge Z. Thus the dipole approximation used above is somewhat marginal for hydrogen, but quite reasonable for heavier species. In our calculations we took abundances for H, He, C, N, and O from standard solar model results, and adopted Cameron abundances for heavy metals (Ne, Mg, Si, S, Fe). The results (see Fig. 2) show that hydrogen accounts for about 1% of the captures, despite its high abundance: free-bound transition are dominated by the heavier metals. Thus the dipole approximation is very well justified.

The importance of the heavier metals results from the fact that the electron's appreciable momentum, $p_e \sim \sqrt{2m_e(\frac{3}{2}T)} \sim 45$ keV, must be absorbed by the bound state, so that compact high $Z$ atomic states containing such high momentum components are favored. The dipole matrix elements depend on the parameter $\eta = Z/a_0 p_e \sim 0.08Z$. For small $\eta$ (thus small $Z$), the transition probabilities vary as $\eta^{2\ell}$ and strongly favor capture into $s$ states. As long as $\eta$ is small, the $n = 1$ state dominates the capture, accounting in hydrogen for 97% of the total at the peak of the free-bound neutrino spectrum. However this is not the case with high $Z$ ions, when $\eta$ becomes comparable to $n$: the $n = 1$ transition in Fe accounts for only $\sim 60\%$ of the total cross section. In our calculations we summed all transitions through $n = 5$, which we found necessary for achieving 0.1% accuracy in the case of Fe.

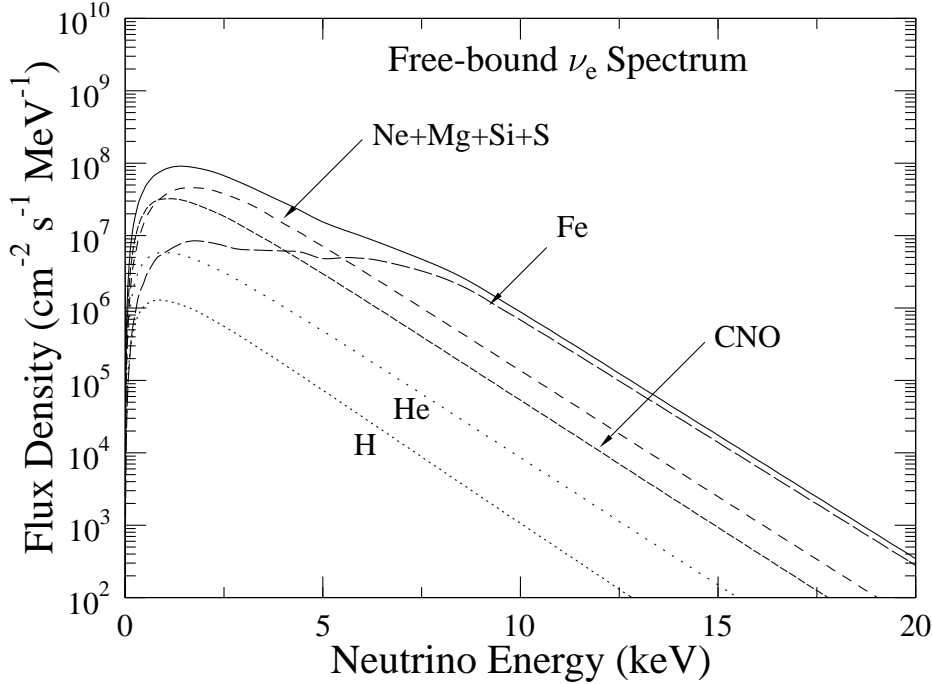

FIG. 2. The total free-bound $\nu_e$ flux density (solid line) and the separate contributions of Fe, Ne+Mg+Si+S, the CNO elements, He, and H are shown as a function of the neutrino energy. Ignoring corrections of order $q_0/m_e$, the $\bar{\nu}_e$ results are identical. The heavy-flavor neutrino flux density is about a factor of 0.46 smaller.

Figure 2 shows that the heavy $\alpha$-stable metals, Ne, Mg, Si, and S, dominate the free-bound rate near the peak of the distribution, but give way to Fe at large $\epsilon_\nu$. This reflects the kinematic advantage of Fe because of its larger atomic binding energy. At very low energies the CNO elements are the largest contributor. One might hope that the thermal solar neutrino flux could be exploited to probe the metallicity of the solar core, testing one of the assumptions of the standard solar model. As the sun is believed to have been highly convective at the onset of the main sequence, and thus homogenized, the standard solar model equates the initial core metal abundances to present-day surface abundances. It would be nice to verify this assumption experimentally by detecting the free-bound neutrino flux. Unfortunately we will see below that the free-bound flux is generally hidden by other, stronger thermal neutrino processes.

The contributions to the terrestrial neutrino flux from solar thermal neutrino processes were obtained by summing the contributions from the five processes described above and depicted in Fig. 1. The rates were folded with the standard solar model density and temperature profiles [11], which were also employed in calculating the electron chemical potential as a function of radius. Integrals were done over the solar core with a sufficient number of Gaussian mesh points to achieve a numerical accuracy of at least 0.1%. The distributions of heavy metals not taken from the standard solar model (Ne, Mg, Si, S, Fe) were assumed to be uniform throughout the sun: no attempt was made to estimate effects due to diffusion.

Figure 3 gives the resulting $\nu_e/\bar{\nu}_e$ and heavy-flavor flux densities. (Note that certain thermal processes, such as bound-bound transitions, will contribute but have not been estimated: as bound-bound transitions are known to be relatively unimportant to the solar opacity, we assumed they would also be small in the weak case. This process can be considered an omitted correction due to incomplete atomic ionization in the solar core.) The figure shows that the simple Compton process dominates the high-energy tails of both the $\nu_e/\bar{\nu}_e$ and heavy-flavor spectra, $\epsilon_\nu \gtrsim 5$ keV. In the heavy neutrino spectrum, the free-bound process dominates below 2 keV. Clearly the only "window" on core metallicity is then a tiny one, involving the enormously difficult task of observing very low energy heavy-flavor neutrinos. In the case of electron neutrinos, the pole contribution to the Compton process dominates below 5 keV. (The pole Compton and plasmon rates are proportional to $g_V^2$ and thus are very much smaller for heavy flavor neutrinos.) One concludes that the entire $\bar{\nu}_e$ spectrum – this is the species most likely to be detected some day – is governed by purely leptonic thermal processes that can be readily calculated from the core temperature distribution and electron density. Thus it is information on the core temperature that is encoded in this flux.

In Fig. 4 we superimpose the spectrum of thermal neutrinos, generated by summing all of the processes discussed here, on a familiar graph [12] of sources contributing to the neutrino flux density at earth. The thermal contributions are significant. The peak of the thermal solar $\nu_e$ (or $\bar{\nu}_e$) flux density occurs at low energies, where the dominate production mechanisms are plasmon decay and the plasmon pole contribution to the Compton process. The peak flux density is $\sim 10^9$/cm$^2$/sec/MeV, a value somewhat less than 1% that achieved at the peak of the pp solar fusion neutrino distribution and about an order of magnitude larger than the terrestrial $\bar{\nu}_e$ peak. The heavy-flavor flux density reaches its maximum at $\sim 2$ keV, achieving a peak value about an order of magnitude below that found for electron neutrinos. It is the only important heavy-flavor contribution in Fig. 4 above cosmic background energies. The thermal $\nu_e/\bar{\nu}_e$ spectrum is the dominate feature above cosmic background energies and below $\sim 5$ keV, where the solar fusion $\nu_e$ and terrestrial $\bar{\nu}_e$ distributions take over. The integrated $\nu_e$ and $\nu_\mu$ fluxes are $2.1 \cdot 10^6$/cm$^2$/sec and $3.0 \cdot 10^5$/cm$^2$/sec, respectively.

In using Fig. 4 it is important to remember that in our evaluation of the Compton pole process, and in our neglect of longitudinal plasmon decay, we have made approximations that should break down near and below $\epsilon_\nu \sim \omega_{pl}$. Thus these contributions to the low-energy results ($\lesssim 0.3$ keV) in Fig. 4 are a naive extrapolation, and should be viewed with caution.

The thermal neutrinos are unimportant to solar evolution as they account for a negligible fraction of the energy loss ($\sim 0.001$%). However, the possibility that they could be exploited to check the solar core temperature is intriguing. The rate of the Compton process varies as $T^7$. Thus a 5% change in the core temperature – a variation that would significantly reduce the $^8$B solar neutrino flux – would be expected to change the thermal neutrino flux by about 30%. The high-energy tail of the distribution would be considerably more sensitive to temperature variations, of course. This cross check on the standard solar model might be interesting because it is independent of nuclear cross sections, on composition gradients, or any other detailed aspects of the model. Another nice property of such a check is that the thermal $\bar{\nu}_e$ flux is left unaffected by the simplest MSW neutrino oscillation scenario [13]: $\bar{\nu}_e$s do not experience a level crossing because the effective mass of the $\bar{\nu}_e$ has the wrong sign, assuming the usual mass hierarchy where the electron type is the lightest flavor. This contrasts with the case of solar fusion $\nu_e$s: in principle, the comparison of the pp, $^7$Be, and $^8$B neutrino fluxes is very sensitive to the core temperature, but in practice the fluxes appear to reflect neutrino oscillations, making the "thermometer" somewhat harder to read. Thus a second thermometer with quite different sensitivity to new physics (both in the MSW mechanism and in $\delta m^2/\epsilon_\nu$) could be quite valuable.

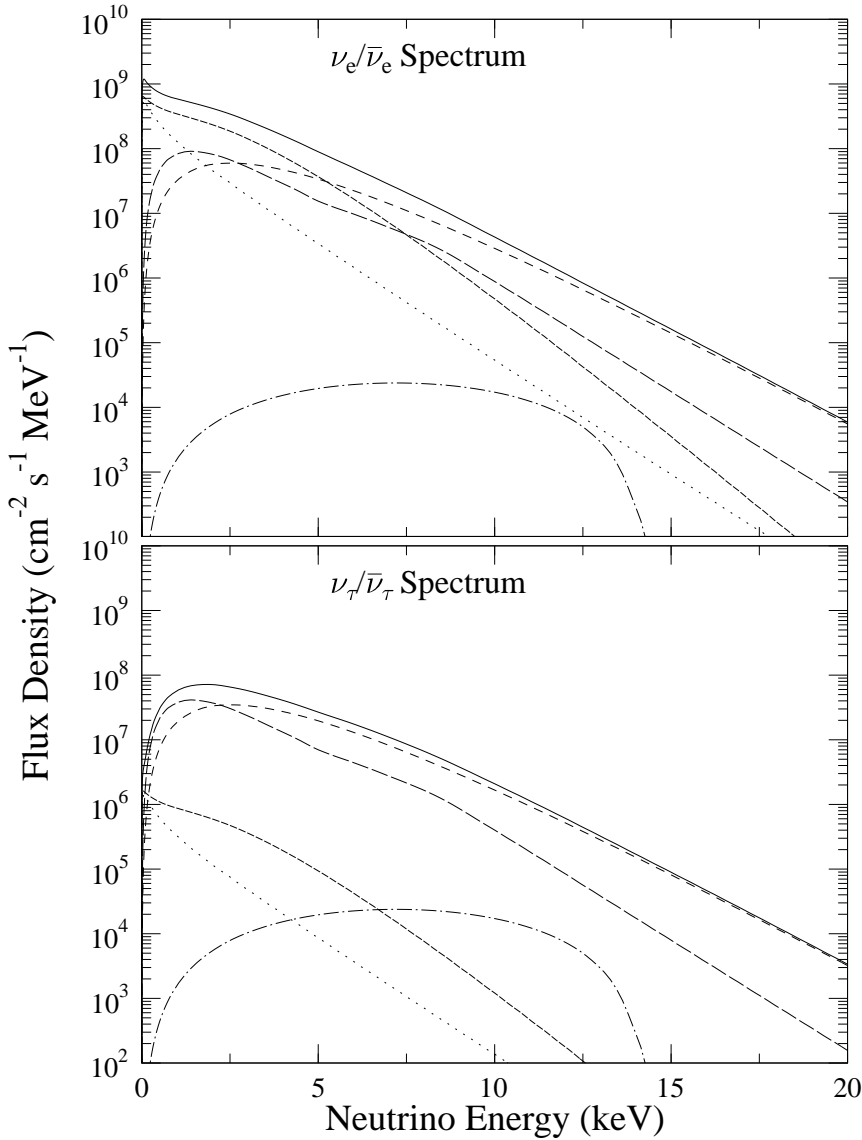

FIG. 3. Neutrino flux densities from free-bound transitions (long dashes), the Compton pair process (intermediate dashes), the pole contribution to the Compton pair process (short dashes), $^{57}$Fe nuclear decay (dashed-dot), and transverse plasmon decay (dotted). Totals (solid line) are for a single flavor.

In Fig. 5 we show where $\bar{\nu}_e$s of different energy are produced within the sun. Clearly the spectrum contains a great deal of information on the temperature *distribution* within the sun.

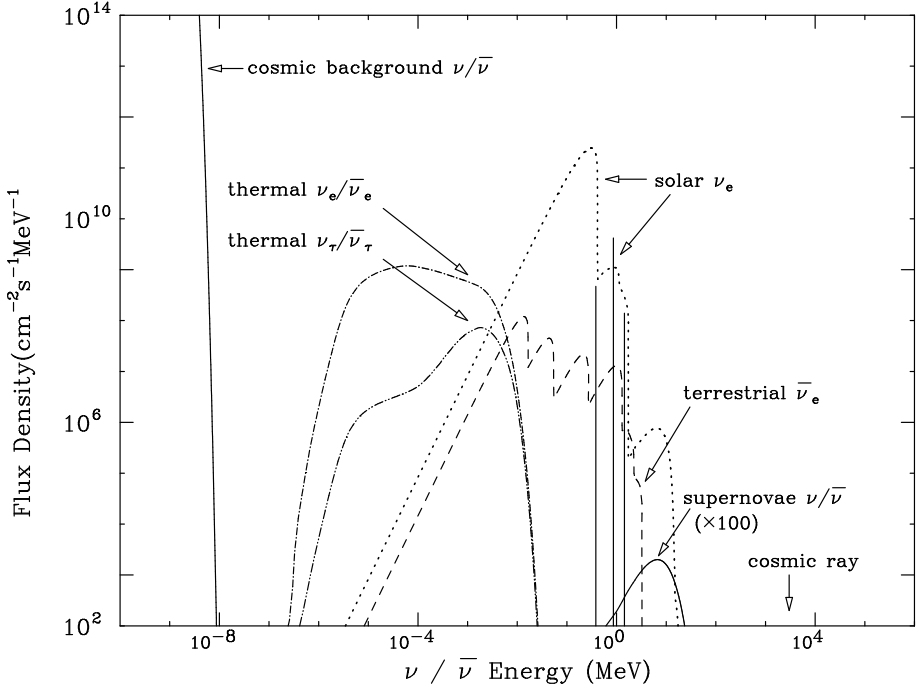

FIG. 4. Natural neutrino sources. The terrestrial $\bar{\nu}_e$ flux and continuous flux of extragalactic supernova neutrinos of all flavors are from Krauss *et al.* [12]. The solar (fusion) $\nu_e$ flux is the standard solar result of Bahcall *et al.* [11]. The thermal solar neutrinos are for a single flavor.

These remarks are made because the most likely opportunity for measuring the thermal neutrino spectrum is a process that depends on flux density, not on total flux, and which samples that flux at a precise energy, the resonant reaction

$$\bar{\nu}_e + e^- + (A, Z) \to (A, Z - 1). \qquad (23)$$

This reaction has been discussed previously in connection with terrestrial $\bar{\nu}_e$ sources [12]. Cross sections can be large in high $Z$ atoms, where the electron overlap with the nucleus is favorable. Because nuclear level widths are very narrow, this process samples the $\bar{\nu}_e$ flux density at a discrete energy. The are several possible candidate transitions with energies between 2 and 20 keV. (One that has been studied in connection with neutrino mass measurements is the decay of long-lived $^{163}$Ho to $^{163}$Dy, which has a positive q-value of less than 3 keV: either a neutrino mass or $\bar{\nu}_e$ inducement of electron capture alters the atomic orbits that participate in the capture.)

The heavy-flavor neutrino flux also contains interesting information: if the existence of this flux were established, it would immediately impose kinematic mass limits of $\sim 1$ keV on the $\nu_\mu$ and $\nu_\tau$. Unfortunately there is no obvious possibility for measuring these species. The problem could well prove as difficult as in the case of the cosmic microwave neutrinos, where existing experimental bounds exceed the expected flux by about 15 orders of magnitude [14].

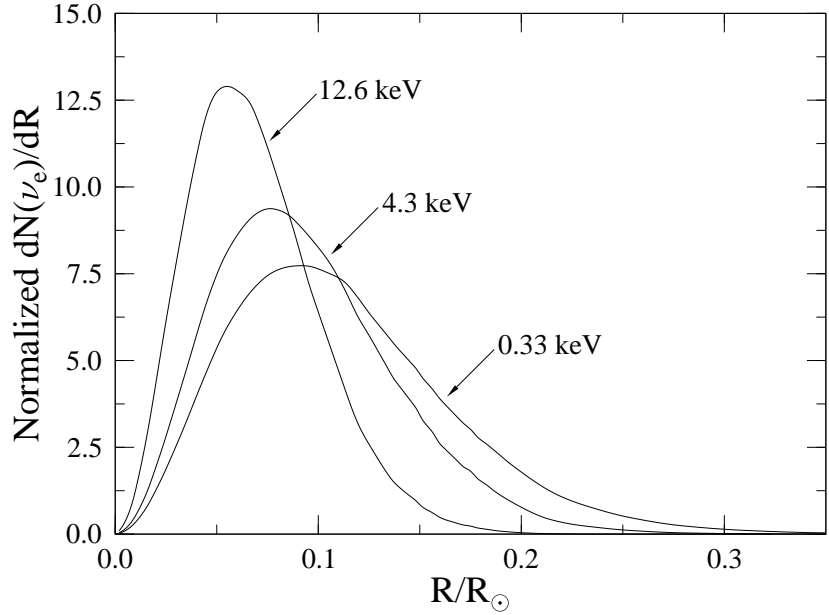

FIG. 5. The radial distribution of neutrino pair production within the sun for three selected energies, 0.33, 4.3, and 12.6 keV. Each distribution, integrated over R, is normalized to unity.

This work was supported in part by the US Department of Energy and by NASA under contract NAGW-2523.
† Present address: JiaChina, Inc., 2785 Lawrenceville Hwy, Suite 106, Dacatur, GA 30033